\documentclass[sigplan,10pt]{acmart}
\renewcommand\footnotetextcopyrightpermission[1]{}
\usepackage{microtype}
\usepackage{enumitem}
\setlist{topsep=2pt,itemsep=0pt,parsep=0pt,partopsep=0pt}

\usepackage{amsmath}

\usepackage{tikz}
\usepackage[labelformat=simple]{subcaption}

\usepackage{booktabs}
\usepackage{multirow}
\usepackage{array}
\usepackage{tabularx}

\usepackage{algorithm}
\usepackage{algpseudocode}

\usepackage{xcolor}
\usepackage{xspace}
\usepackage{enumitem}

\hypersetup{
    colorlinks=true,
    linkcolor=blue!90!red,
    citecolor=magenta
}

\newenvironment{myitemize}
  {\begin{list}{\labelitemi}{
      \itemsep=1pt
      \topsep=2pt
      \parsep=0pt
      \partopsep=0pt
      \leftmargin=1.4em
  }}
  {\end{list}}

\newcommand{\heading}[1]{%
    \vspace{1ex}%
    \noindent
    \textbf{#1}%
}

\setenumerate[1]{
    itemsep=0pt,
    partopsep=0pt,
    parsep=\parskip,
    topsep=5pt
}

\setitemize[1]{
    itemsep=0pt,
    partopsep=0pt,
    parsep=\parskip,
    topsep=5pt
}

\newcommand{\sys}{\textsc{Vitamin-E}\xspace}

\begin{document}

\date{}

% \title{Determinism-Preserving GPU Spatial Sharing: Late-Binding under the Parallel-Structure Invariant}

\title{Determinism-Preserving GPU Spatial Sharing with \sys}

\author{Zhenyuan Yang}
\authornote{Equal contribution}
\affiliation{
  \institution{Key Laboratory of System Software (Chinese Academy of Sciences)}
  \institution{Institute of Software, Chinese Academy of Sciences}
  \institution{University of Chinese Academy of Sciences}
  \country{}
}

\author{Wenxin Zheng}
\authornotemark[1]
\affiliation{
  \institution{Shanghai Jiao Tong University}
  \country{}
}

\author{Mingyu Li}
\affiliation{
  \institution{Key Laboratory of System Software (Chinese Academy of Sciences)}
  \institution{Institute of Software, Chinese Academy of Sciences}
  \country{}
}

\author{Haibo Chen}
\affiliation{
  \institution{Key Laboratory of System Software (Chinese Academy of Sciences)}
  \institution{Institute of Software, Chinese Academy of Sciences}
  \institution{Shanghai Jiao Tong University}
  \country{}
}

\begin{abstract}
GPU sharing faces a determinism--utilization tradeoff: fixed bindings can strand capacity as demand fluctuates, while resource-driven kernel reshaping improves utilization by altering a launch's parallel structure, potentially changing output bits.
We rethink modern GPU scheduling and observe that it decouples logical structure from physical width: one unmodified launch spans a family of widths through changes in block placement and wave count.
From this observation, we derive the parallel-structure invariant: for fixed-structure deterministic workloads, keeping each launch immutable makes its output bits independent of physical width.

Guided by this invariant, \sys late-binds immutable launches to pooled physical contexts, preserving bitwise equality across allocations, whereas resource-driven reshaping can alter the selected token under temperature-zero greedy decoding.
Across all workload--baseline comparisons, \sys achieves up to 3.50$\times$ the aggregate normalized LLM training throughput, 62.5\% lower inference p99 latency, and 1.43$\times$ the background-training throughput.
With the same mechanism, \textsc{TPOT-First} reduces TPOT SLO violations by up to 46.1\% over \textsc{Throughput-Oriented} on three serving workloads, demonstrating mechanism effectiveness and policy flexibility.
\end{abstract}

\maketitle
\pagestyle{plain}

\section{Introduction}
Machine learning (ML) workloads expose rapidly changing GPU demand and exploitable parallelism across execution phases, request arrivals, batch sizes, and input characteristics~\cite{jeon2019analysis,shukla2022singularity,alpaserve,mixedfusion,fastermoe}.
Cloud operators often provision for peak demand to satisfy service-level objectives (SLOs), leaving substantial compute capacity unused during less-parallel phases~\cite{Chen_Latency}.
Spatial sharing~\cite{pagoda,sgdrc,yu2020salus,strati2024orion,coppock2025lithos,Huang2026ShareNK} can reclaim this capacity by colocating workloads with complementary demands, provided that resource allocation can follow short-lived demand changes at kernel-launch granularity.

Determinism is important for reproducibility, debugging, regression testing, and stable ML evaluation~\cite{doradd,deepseekai2026deepseekv4highlyefficientmilliontoken,song-etal-2025-good,zhang2026deterministicinferencetensorparallel}, as discussed in~\autoref{sec:determinism_matter}.
Resource adaptation should therefore not introduce output variation into an otherwise deterministic execution.
For a deterministic workload, varying the physical allocation may affect execution time but should not alter its parallel structure.
If that structure depends on transient resource state, runs differing only in physical allocation can produce different results as co-tenant activity changes, thereby introducing additional nondeterminism.

\begin{figure}[t]
\centering
    \includegraphics[width=\columnwidth]{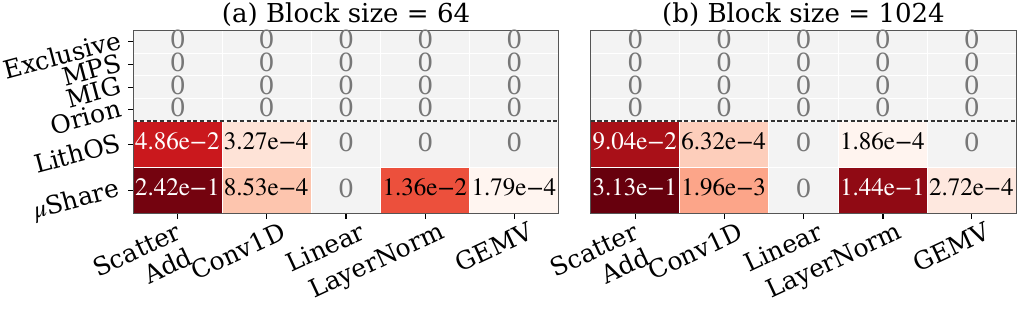}
    \caption{Maximum deviation from exclusive-GPU execution over 1,000 matched-input trials. Fixed-structure mechanisms are bitwise identical. LithOS and $\mu$Share show drift.}
    \label{fig:exist-system}
\end{figure}

Existing spatial-sharing mechanisms fail to provide both launch-granular adaptation and determinism.
NVIDIA Multi-Process Service (MPS), Multi-Instance GPU (MIG), and Green Contexts preserve kernel structures but bind resources to a process, GPU instance, or execution context across many launches~\cite{nvidia_mps_guide,nvidia_mig_guide,NVIDIA2025Green}.
These fixed bindings can strand capacity as demand changes.
Kernel-level systems adapt more frequently by changing block sizes, launch geometry, or work decomposition according to available resources~\cite{kernelet,coppock2025lithos,Huang2026ShareNK,hu2026hummingbirdsloorientedgpupreemption}.
Such reshaping can alter floating-point accumulation order in LayerNorm, GEMV, and related operators.
As~\autoref{fig:exist-system} shows, reshaping~\cite{coppock2025lithos,Huang2026ShareNK} produces nonzero output drift across matched inputs, whereas fixed-structure systems remain bitwise identical.
This creates a tradeoff between coarse structure-preserving bindings and fine-grained structure-changing adaptation.

We rethink modern GPU scheduling and observe that it already separates logical structure from physical allocation.
A submitted launch specifies its logical blocks and work decomposition, while the hardware scheduler maps those blocks onto the eligible SMs.
The same unmodified launch can therefore execute at different physical widths, with only block placement and wave count changing.
This yields the \emph{parallel-structure invariant}: physical allocation may vary while the kernel binary, launch geometry, work decomposition, and required ordering remain fixed.
For a fixed-structure deterministic workload, a resource manager can thus late-bind each immutable launch to an allocation selected from current availability, gaining flexibility without kernel reshaping.

We present \sys, a GPU runtime that realizes spatial sharing under this invariant.
A virtual context (\texttt{vCtx}) owns immutable operation descriptors and application-required dependencies.
A physical context (\texttt{pCtx}) represents one replaceable allocation of SMs or compute units.
Once an operation becomes logically ready, a policy-driven dispatcher selects an eligible \texttt{pCtx} and submits the original descriptor unchanged.
This separation allows successive launches to use different physical allocations while their parallel structure and required ordering remain fixed.

Realizing this separation is non-trivial and brings three challenges.
First, redirecting operations across backend contexts can violate stream- and event-defined happens-before order; \sys therefore encodes stream, event, and default-stream dependencies as a logical happens-before DAG in the \texttt{vCtx}, preserving the application-defined partial order across changing \texttt{pCtxs}.
Second, efficient adaptation must avoid both dispatch-path context creation and exhaustive pre-creation; \sys addresses this with a bounded hierarchical \texttt{pCtx} pool supporting non-overlapping leases, reservation, and lookahead-driven reclamation.
Third, policy flexibility must not expose logical state; a narrow interface lets policies select ready operations, eligible allocations, and preparation actions, while the dispatcher validates proposed binding.

We evaluate \sys along four dimensions: determinism, colocated training, mixed training--inference, and policy flexibility.
For fixed-structure deterministic workloads, \sys preserves bitwise equality across physical allocations in operator and decoding experiments, whereas reshaping can alter the selected token under temperature-zero greedy decoding.
Across colocated LLM training workloads, \sys achieves 1.38--1.63$\times$ the geometric-mean aggregate normalized throughput of the baselines.
For mixed training--inference workloads, using geometric means across workloads, \sys reduces inference p99 latency by 8.1--58.4\% and achieves 1.05--1.36$\times$ the background-training throughput of the baselines.
Finally, \textsc{TPOT-First} reduces time per output token (TPOT) SLO violations by 21.2--46.1\% relative to the \textsc{Throughput-Oriented} policy, demonstrating the flexibility of \sys's policy interface.

\noindent This paper makes the following contributions:

\begin{myitemize}
\item We identify the tradeoff between determinism and utilization in GPU spatial sharing and derive the parallel-structure invariant from hardware's separation of logical structure and physical width.

\item We design and implement \sys, which combines dependency-preserving virtual contexts, a hierarchical physical-context pool, and validated late binding to adapt each ready launch without changing its descriptor.

\item We evaluate \sys on operators, large language model (LLM) decoding, colocated training, mixed workloads, and trace-driven serving.
It preserves bitwise equality for fixed-structure deterministic workloads while improving throughput, tail latency, and TPOT SLO attainment.
\end{myitemize}

\section{Motivation}
\label{sec:motivation}

\subsection{Why GPU Determinism Matters}
\label{sec:determinism_matter}

GPU determinism means that the same application-defined computation in the same environment produces the same output bits.
GPU resource sharing should preserve this property when changing the physical resources used by an operation. Otherwise, outputs may depend on transient allocation decisions and unrelated co-tenant activity.

\begin{figure}[t]
    \centering
    \includegraphics[width=\columnwidth]{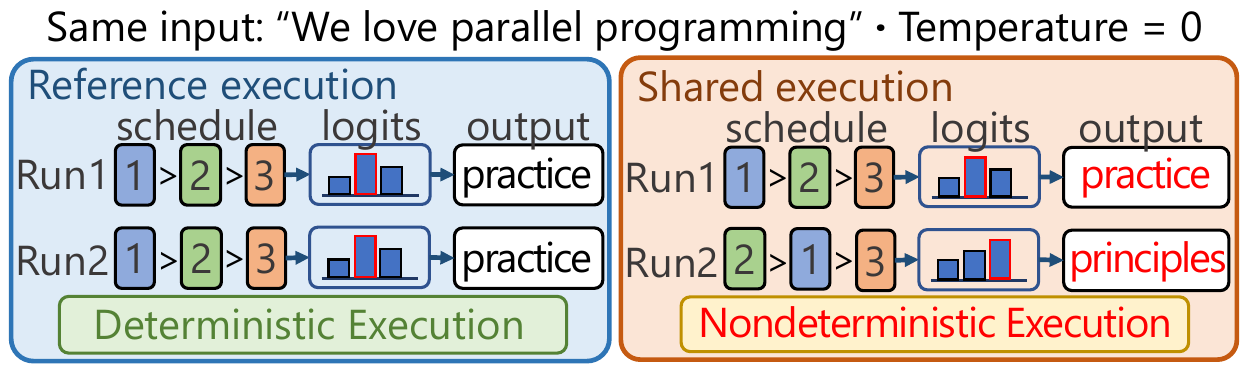}
    \caption{A numerical perturbation can change the greedy-decoding argmax at a high-entropy step and cause subsequent tokens to diverge.}
    \label{fig:token-divergence}
\end{figure}

\heading{Consequences of Nondeterminism.}
Nondeterministic outputs hinder failure reproduction, regression testing, and numerical validation because rerunning the same binary and input may produce different results, while schedule-dependent failures may disappear~\cite{doradd,gond2026llm42enablingdeterminismllm,deepseekai2026deepseekv4highlyefficientmilliontoken}.
In ML pipelines, run-to-run variation can cause evaluation drift and inconsistency between rollout and training engines despite fixed models and data~\cite{song-etal-2025-good, zhang2026deterministicinferencetensorparallel, yao2025offpolicy}.
During autoregressive inference, a small logit perturbation can flip the argmax at a high-entropy step and cause subsequent decoding to diverge~\cite{he2025nondeterminism}, as illustrated in~\autoref{fig:token-divergence}.
Iterative and numerically sensitive HPC computations face similar risks when perturbations accumulate and invalidate reference comparisons~\cite{shanmugavelu2024impactsfloatingpointnonassociativityreproducibility}.

These consequences motivate bitwise equality as the resource-control criterion.
Resource adaptation may change latency, throughput, and execution overlap, but it should not change the output of an otherwise deterministic application.

\subsection{Determinism Model}

\heading{Fixed-Structure Determinism.}
Consider an execution of kernel launches indexed by $k$.
For each launch, let $B_k$ denote its kernel binary and $C_k$ its submitted parallel structure, including its launch geometry and kernel-defined work decomposition.
Let $X$ denote the input and initial device-visible state, and let $H$ denote the happens-before order required by application streams, events, and synchronization.
The execution is \emph{fixed-structure deterministic} if repeated runs in the same hardware and software environment, with identical $X$, the same $B_k$ and $C_k$ for every launch, and the same $H$, produce bitwise-identical outputs.

\heading{Determinism under Resource Adaptation.}
A spatial-sharing runtime may select a different eligible physical allocation $P_k$ for each kernel launch $k$.
Resource adaptation preserves fixed-structure determinism if any two runs that differ only in these per-launch allocation choices preserve every $B_k$ and $C_k$, as well as $H$, and produce bitwise-identical final outputs.
Physical allocation may therefore change placement, execution overlap, and completion time, but not the application-visible result.

This contract applies only to eligible physical-allocation changes within a fixed hardware and software environment.
It makes no claim about runs that change the input, application-defined execution, or environment.

\subsection{Sources of Nondeterminism}

\heading{Application and Environment Sources.}
Explicit random inputs, data races, unsynchronized accesses, and clock-dependent code can make an application nondeterministic even when its parallel structure and physical allocation are fixed.
Changes in hardware architecture, compiler, library, algorithm, or numerical precision can also change output bits across environments.
These sources lie outside the resource-control contract.
The contract applies only to executions that are already fixed-structure deterministic and asks whether resource adaptation introduces additional output variation.

\begin{figure}[!t]
    \centering
    \begin{subfigure}[!t]{0.58\columnwidth}
        \centering
        \includegraphics[width=\linewidth]{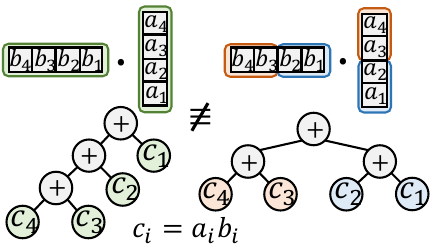}
        \caption{Different reduction orders.}
        \label{fig:reshaping-tree}
    \end{subfigure}
    \hfill
    \begin{subfigure}[!t]{0.41\columnwidth}
        \centering
        \includegraphics[width=\linewidth]
        {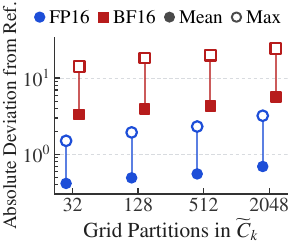}
        \caption{Numerical deviations.}
        \label{fig:reshaping-deviation}
    \end{subfigure}

    \caption{Resource-driven reshaping repartitions a reduction kernel, changing its reduction tree (a) and yielding nonzero FP16/BF16 deviations from the fixed-structure reference (b).}
    \label{fig:reshaping-determinism}
\end{figure}

\heading{Resource-Dependent Parallel Structure.}
ML workloads create pressure for fine-grained resource adaptation because their GPU demand changes across compute, communication, transfer, LLM prefill, and decode phases~\cite{jeon2019analysis, shukla2022singularity, alpaserve, mixedfusion, fastermoe, peng2019bytescheduler, TicTac, agrawal2024taming, zhu2025nanoflow, kamath2025pod}.
Bursty arrivals, variable batch sizes, and input-dependent computation add further variation~\cite{SHEPHERD,inferline}.
Peak provisioning therefore leaves compute capacity idle whenever instantaneous demand falls below the peak.

Temporal sharing multiplexes workloads over time, incurring context-switch overhead and preventing workloads with complementary resource demands from running in parallel~\cite{effisha, xiao2018gandiva,xiao2020antman,wu2023transparent,clock,Chimera,ausavarungnirun2018mask,bai2020pipeswitch}.
Spatial sharing instead allows kernels to run in parallel on separate compute allocations.
MPS, MIG, and Green Contexts preserve submitted kernel structure, but their process-, instance-, or context-level allocations span many launches or require coarse reconfiguration or redirection among stateful execution objects~\cite{nvidia_mps_guide, nvidia_mig_guide, NVIDIA2025Green}.

Kernel-level systems obtain finer-grained adaptation by changing block sizes, launch geometry, or work decomposition~\cite{kernelet, coppock2025lithos, Huang2026ShareNK, hu2026hummingbirdsloorientedgpupreemption}. For example, $\mu$Share changes block sizes, while LithOS decomposes kernels into execution atoms~\cite{Huang2026ShareNK, coppock2025lithos}. Such systems derive the structure that actually executes from the submitted structure and the resource state observed at launch time: $\widetilde C_k=g(C_k,R_k)$, where $R_k$ denotes the transient resource state and $\widetilde C_k$ the resulting executed structure. Two otherwise identical runs may encounter different co-tenant activity and therefore obtain different values of $\widetilde C_k$, even though the application submitted the same $C_k$.

IEEE~754 arithmetic rounds each operation to finite precision~\cite{goldberg1991every}.
Let $\oplus$ denote addition in the working precision.
Two valid parenthesizations of the same reduction may produce different bit patterns: $\bigl((c_1 \oplus c_2) \oplus c_3\bigr) \oplus c_4 \neq (c_1 \oplus c_2) \oplus (c_3 \oplus c_4)$. In a parallel reduction, the executed parallel structure $\widetilde C_k$ determines how partial results are formed and combined. Changing $\widetilde C_k$ can therefore change the reduction tree and floating-point operation order, as~\autoref{fig:reshaping-determinism}\subref{fig:reshaping-tree} illustrates.
Parallel structure is thus part of the effective numerical computation for reductions, Split-K GEMMs, and similar kernels~\cite{spmm}. \autoref{fig:reshaping-determinism}\subref{fig:reshaping-deviation} demonstrates this effect for a floating-point reduction. Different grid partitions produce nonzero deviations from the fixed-structure result, with larger deviations in BF16 than in FP16 for this kernel.

Not every change in parallel structure changes an output.
However, because $\widetilde C_k$ depends on the transient resource state $R_k$, two runs with the same input, submitted kernel, and application-required ordering may execute different structures.
Such runs no longer differ only in their physical allocations: the resource controller has changed the effective numerical computation.
Resource-dependent reshaping therefore violates the fixed-structure premise of the resource-adaptation contract and can introduce output variation into an otherwise fixed-structure deterministic execution.
\section{The Parallel-Structure Invariant}
\label{sec:invariant}

We first show that resource adaptation threatens determinism when it changes the executed parallel structure, whereas physical allocation alone need not alter the submitted computation.
We then derive an invariant from hardware block scheduling that separates the two.

\subsection{Structure Invariance under Physical Allocation}

For each launch $k$, the GPU scheduler maps the logical blocks specified by the submitted structure $C_k$ onto the selected physical allocation $P_k$, issuing one or more waves as needed.
Changing $P_k$ may alter block placement, the number of execution waves, execution overlap, and completion time.
It does not inherently require changing the kernel binary $B_k$ or the submitted parallel structure $C_k$.

This separation yields the \emph{parallel-structure invariant}: $C_k$ remains equal to the submitted structure as $P_k$ changes.
Hardware scheduling already preserves this invariant when executing a fixed grid on different eligible physical allocations.
If the runtime also preserves $B_k$ for every launch and the application-required happens-before order $H$, two runs that differ only in their choices of $P_k$ remain within the resource-adaptation contract.
A fixed-structure deterministic execution therefore produces the same output bits in both runs.

\subsection{Why Existing Controls Lack This Invariant}

Existing mechanisms couple physical allocation to logical execution in two different ways:

\begin{myitemize}
\item \textbf{Execution-container coupling.}
Process-, instance-, and context-level mechanisms preserve $B_k$ and $C_k$, but bind $P_k$ to a long-lived execution object that also carries streams, events, or other execution state.
Changing $P_k$ across launches therefore requires reconfiguring that object or redirecting operations among multiple stateful objects while reconstructing their dependencies.

\item \textbf{Kernel-structure coupling.}
Kernel-level systems obtain fine-grained resource control by transforming the kernel or its submitted parallel structure.
The resulting executed structure $\widetilde C_k$ depends on transient resource state rather than remaining equal to $C_k$, violating the parallel-structure invariant.
\end{myitemize}

The first coupling limits adaptation granularity.
The second allows the resource controller to change the effective computation.
A suitable resource-control abstraction must separate the parallel structure from the physical allocation.

\subsection{Desired Goals for Invariant-Preserving Control}

We pursue three goals for invariant-preserving resource adaptation:

\begin{myitemize}
\item \textbf{Immutable logical execution.}
Resource adaptation must preserve every submitted kernel binary and parallel structure, together with the application-required happens-before order.

\item \textbf{Efficient launch-granular late binding.}
Each ready launch should be bound to an eligible physical allocation without migrating application-visible state or reconfiguring a stateful object on the dispatch path.

\item \textbf{Enforced separation.}
Scheduling policies choose among ready operations and eligible physical allocations, while the runtime retains immutable descriptors and dependency state and validates each proposed binding.
\end{myitemize}

Together, these goals enable validated late binding between protected logical execution and adaptable physical allocation.

\subsection{Hardware Primitives and the Software Gap}

Current GPU platforms can represent physical allocations without kernel reshaping.
NVIDIA Green Contexts associate execution contexts with provisioned SMs and work queues, while AMD CU-masked queues associate queues with selected compute units~\cite{NVIDIA2025Green,amd_rocm_cu_mask}.
A launch submitted through either object is restricted to its associated resources.

However, physical allocation remains tied to an execution object.
Rebinding successive launches requires selecting among objects and reconstructing dependencies expressed through the application's original streams and events, thereby incurring overhead or risking ordering violations even with unchanged kernel descriptors.
What is missing is an indirection layer that owns logical descriptors and dependencies independently of physical objects and efficiently late-binds each ready operation to an eligible allocation.
\section{\sys Design}
\label{sec:design}

\subsection{Overview}

\begin{figure}[!t]
    \centering
    \includegraphics[width=\columnwidth]{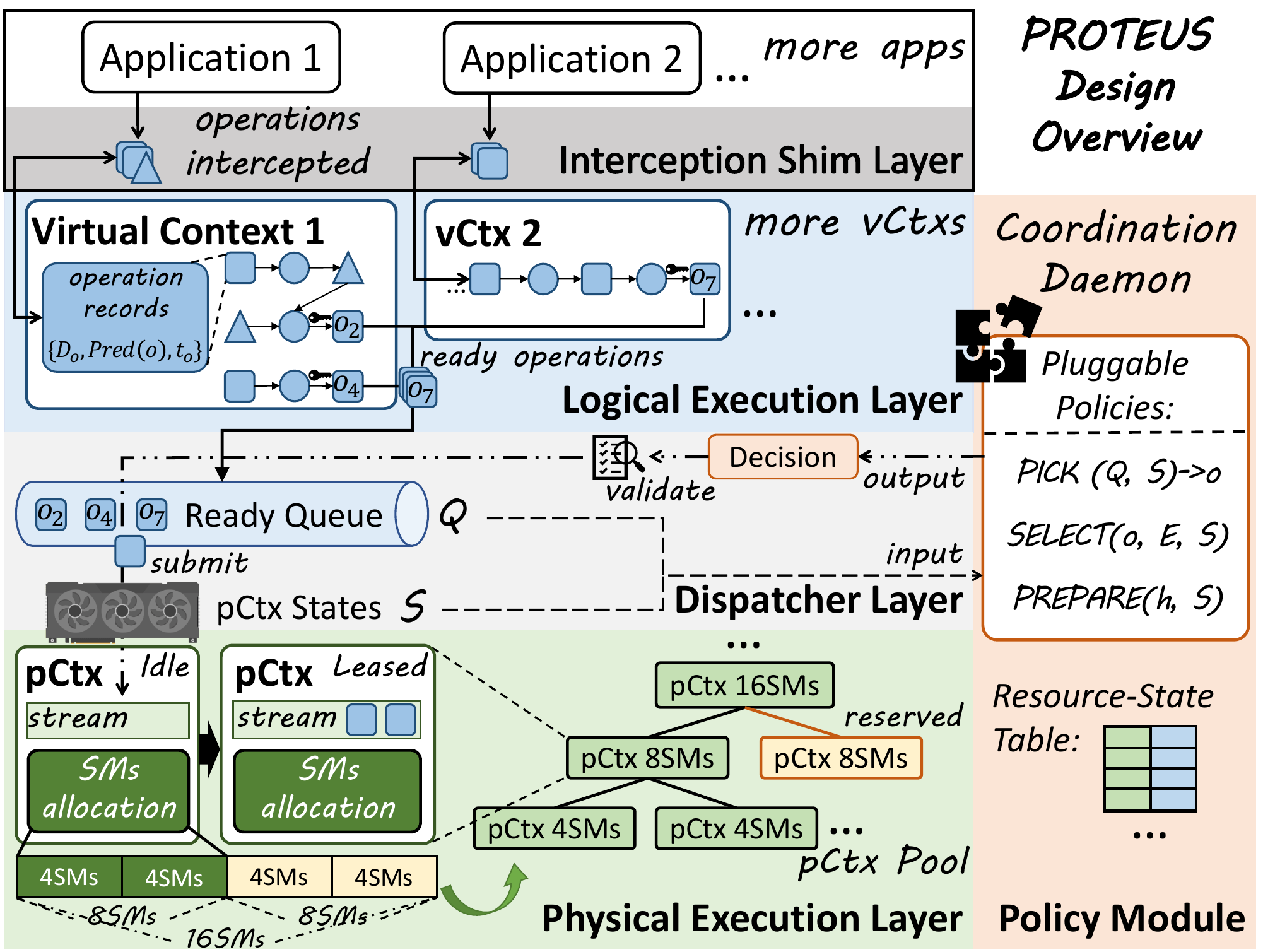}
    \vspace{-0.5em}
    \caption{The overall architecture of \sys, which keeps logical descriptors and dependencies in virtual contexts and binds ready operations to replaceable physical contexts.}
    \label{fig:design-overview}
\end{figure}

\autoref{fig:design-overview} shows three components that enforce the parallel-structure invariant:

\begin{myitemize}
\item \textbf{Virtual context (\texttt{vCtx}).}
A \texttt{vCtx} owns the logical execution state, including application-visible streams, events, synchronization state, immutable operation descriptors, and dependencies.
It has no fixed allocation (\autoref{sec:vctx}).

\item \textbf{Physical context (\texttt{pCtx}).}
A \texttt{pCtx} represents one replaceable physical allocation of SMs or compute units.
It contains only the backend streams, work-queue configuration, and runtime state needed to use that binding (\autoref{sec:pctx}).

\item \textbf{Policy-driven dispatch layer.}
The policy chooses a ready operation and an eligible \texttt{pCtx}.
The dispatcher validates the binding and submits the immutable descriptor through a backend stream (\autoref{sec:dispatch}).
\end{myitemize}

For each intercepted operation, the \texttt{vCtx} records an immutable descriptor and predecessor tokens.
The operation enters the ready queue after those tokens are satisfied.
The policy then proposes a \texttt{pCtx}, and the dispatcher submits the original descriptor only after checking descriptor compatibility, dependency readiness, and lease state.
Completion satisfies the operation's logical token and may release its dependents.
Successive operations from one logical stream may use different \texttt{pCtxs}, while their descriptors and application-required order remain fixed.

This division separates mechanism from policy through a narrow control boundary.
Policies control physical assignment, ready-operation ordering, reservation, and resource preparation, while the runtime retains control over immutable descriptors, logical readiness, and binding eligibility.
Every accepted decision therefore remains within the physical freedom allowed by the parallel-structure invariant.

\subsection{Virtual Contexts: Immutable Logical State}
\label{sec:vctx}

A \texttt{vCtx} stores the descriptors and dependencies that must remain invariant across physical allocations.

\begin{figure}[!t]
    \centering
    \includegraphics[
        width=\columnwidth
    ]{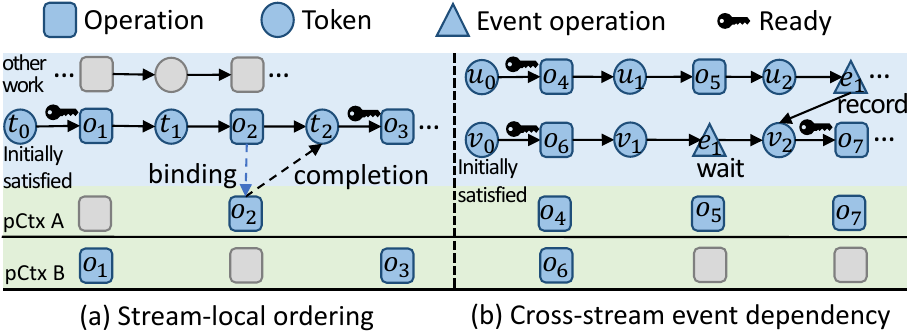}
    \caption{
        Token-based logical dependencies in \sys.
    }
    \label{fig:logical-dependencies}
    \vspace{-0.5em}
\end{figure}

\heading{Token-Based Logical Dependencies.}
\sys represents the completion of each logical operation with a token that is satisfied when its physical execution completes. 
An operation records the tokens of its logical predecessors and becomes eligible for dispatch only after all of them are satisfied. 
For each application-visible stream, the \texttt{vCtx} stores the token of its most recently submitted operation as the stream tail. 
A new operation depends on the current tail, receives a new token, and updates the tail.
An empty stream begins with an already-satisfied token.
This chain preserves stream-local order across changing physical streams or \texttt{pCtxs} (\autoref{fig:logical-dependencies}(a)). 
Events connect these per-stream chains. Each record creates an immutable generation bound to the recording stream's completion point. 
A wait captures the event generation visible when it is issued and depends on that generation's token.
Subsequent records of the same event cannot retarget the wait to a later completion point (\autoref{fig:logical-dependencies}(b)).

Together, stream-order, event-induced, and default-stream dependencies define the application-level happens-before graph:
\[
H =
\bigl(
\mathcal{O},
E_{\mathrm{stream}}
\cup E_{\mathrm{event}}
\cup E_{\mathrm{default}}
\bigr),
\ 
o_i \prec_H o_j
\iff
o_i \leadsto_H o_j ,
\]
where $E_{\mathrm{default}}$ captures the applicable default-stream semantics and $o_i \leadsto_H o_j$ denotes directed reachability in $H$.
For a well-formed execution, $H$ is acyclic. 
By dispatching an operation only after its predecessor tokens are satisfied, \sys preserves $\prec_H$ without logically ordering incomparable operations. 
Any serialization among them is a physical scheduling outcome and adds no edge to $H$.

\heading{Synchronization and Queries.}
Event synchronization waits on the selected generation's token, stream synchronization waits on the stream tail, and device-wide synchronization waits on the current stream tails. 
The corresponding queries inspect the same logical completion state without blocking, independently of any physical stream or \texttt{pCtx}.

\heading{Immutable Operation Descriptors.}
For each kernel launch $k$, the \texttt{vCtx} constructs an immutable descriptor $D_k$ containing the kernel entry, launch-time argument values, grid and block dimensions, dynamic shared-memory size, and launch attributes. 
The edges in $H$ encode its stream ordering and event dependencies. 
\sys copies the argument values required by the launch interface when constructing $D_k$. 
For pointer arguments, it copies only the pointer values.
The referenced device-memory contents remain under application control.
Synchronization of those contents remains the application's responsibility.
Scheduling never modifies $D_k$. 
A \texttt{pCtx} is eligible only if it can execute the descriptor unchanged. 
Other asynchronous operations, including memory copies and memset operations, are represented similarly so that their ordering relationships are preserved.

\heading{Invariant Enforcement.}
The immutable descriptors and application-required order $\prec_H$ define the logical execution protected by \sys.
The scheduler can choose only a ready operation and an eligible \texttt{pCtx}.
It cannot modify a descriptor or dispatch an operation before its required predecessors complete.
For a fixed-structure deterministic workload, accepted allocation sequences preserve $\{B_k\}$, $\{C_k\}$, and $H$.
They therefore satisfy the bitwise-equality contract in~\autoref{sec:motivation}.

\heading{Contract Boundary.}
The guarantee is relative to fixed-structure execution in the same hardware and software environment.
The contract excludes nondeterminism already present in the application.
It also excludes reproducibility across environments.
\autoref{sec:motivation} defines this scope.

\subsection{Physical Contexts: Replaceable Allocations}
\label{sec:pctx}

A \texttt{pCtx} is the backend realization of the variable physical allocation $P_k$.
In the CUDA backend, it comprises an NVIDIA Green Context plus associated backend streams and work-queue state~\cite{NVIDIA2025Green}.
Because it contains no application-visible state for streams, events, descriptors, or dependencies, a \texttt{pCtx} can serve operations from different \texttt{vCtxs} over time without transferring logical state.

\begin{figure}[t]
    \centering
    \includegraphics[width=\columnwidth]{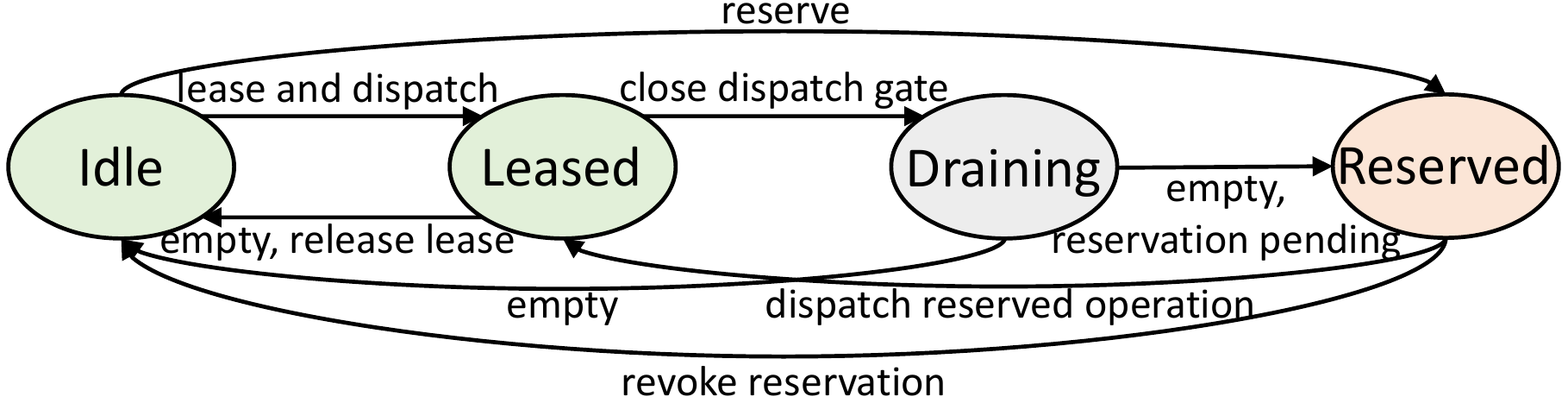}
    \caption{State transitions of a \texttt{pCtx}.}
    \label{fig:pctx-lifecycle}
    \vspace{-1em}
\end{figure}

\heading{\texttt{pCtx} Lifecycle.}
A \texttt{pCtx} can be \emph{idle}, \emph{leased}, \emph{draining}, or \emph{reserved}. 
An idle \texttt{pCtx} has no outstanding work or active resource commitment. 
A leased \texttt{pCtx} holds a lease on its SM allocation and admits operations through an open dispatch gate. 
Closing the gate moves it to draining.
Submitted work continues and no new operations are admitted.
A reserved \texttt{pCtx} is held for a pending operation and is unavailable to unrelated work. 
\autoref{fig:pctx-lifecycle} summarizes the transitions among these states.

\heading{Hierarchical \texttt{pCtx} Pool.}
To keep context and stream initialization off the latency-critical dispatch path, \sys pre-creates a bounded hierarchical \texttt{pCtx} pool inspired by buddy allocation~\cite{knowlton1965fast}.
Leaves represent minimum allocation units, internal \texttt{pCtxs} cover descendant unions, and each node points to a pre-created remainder \texttt{pCtx} spanning the remaining SMs.
For example, an 8-SM node on a 132-SM GPU points to a complementary 124-SM \texttt{pCtx}.
This exposes multiple sizes, including residual allocations, without enumerating arbitrary SM subsets.
Lease conflicts follow directly from the hierarchy: leasing a node excludes its ancestors and descendants, while leasing its remainder excludes the node's ancestors and sibling branches along the path to the root.
Conflicts are therefore identified by hierarchy traversal rather than pairwise \texttt{pCtx} scans.
A \texttt{pCtx} becomes available after its conflicting leases are released.
Unlike conventional buddy allocation, \sys neither creates nor merges resource entities at runtime. The hierarchy only organizes pre-created alternatives, remainder links, and lease constraints.

\begin{figure}[t]
    \centering
    \includegraphics[width=\columnwidth]{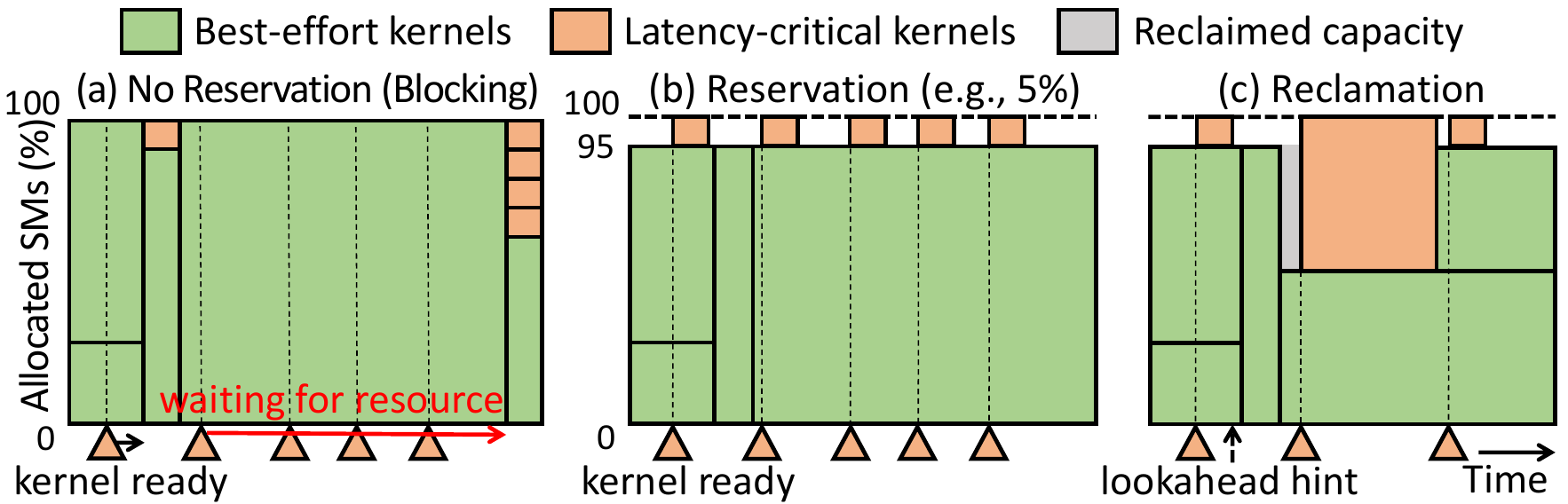}
    \caption{Policy-directed reservation and lookahead-driven reclamation. Reservation avoids resource waiting for recurring latency-critical kernels, while lookahead reclaims a wider allocation before an anticipated kernel becomes ready.}
    \label{fig:reservation-reclamation}
\end{figure}

\heading{Policy-Directed Reservation.}
The physical-context pool allows a policy to reserve a configurable \texttt{pCtx} for latency-critical operations without changing their descriptors.
The reservation excludes its SMs and all overlapping \texttt{pCtxs} from unrelated leases, and its size is chosen based on workload demand, backend allocation granularity, and offline performance profiles. 
As~\autoref{fig:reservation-reclamation}(a)--(b) illustrates, a reserved allocation allows recurring latency-critical operations to dispatch without waiting for overlapping best-effort work to drain. 
Prior work shows that small-batch inference can often approach its minimum latency using only a modest fraction of GPU compute capacity~\cite{choi2022serving,sgdrc}, leaving the remaining resources available to best-effort work.

\heading{Lookahead-Driven Reclamation.}
When an anticipated operation requires a wider allocation, \sys can begin reclaiming the resources before the operation becomes ready. 
Lookahead hints come either from predecessor operations already recorded in a \texttt{vCtx}, such as a memory transfer preceding a decode, or from advisories or configurations specifying resource demand, ready time, or deadline.
As shown in~\autoref{fig:reservation-reclamation}(c), the policy closes the dispatch gates of overlapping leased \texttt{pCtxs}, allowing outstanding work to drain while preventing new admissions.
This overlaps reclamation with predecessor execution, data transfer, or host preparation, hiding much of its latency.
After conflicts drain, \sys reserves the allocation for the anticipated operation.
If reclamation does not complete in time, the policy selects a feasible available allocation or defers dispatch.

\subsection{Dependency-Preserving Dynamic Dispatch}
\label{sec:dispatch}

The dispatch layer maps a fixed parallel structure to a physical allocation for each launch.
It considers only logically ready operations and preserves the operation descriptor and application-required order $\prec_H$ for every dispatch.

\heading{Readiness Before Binding.}
A naive design may bind an operation before its predecessors complete and encode the unresolved dependencies as backend waits. 
Such early binding may occupy a scarce \texttt{pCtx} while the operation cannot execute, and the assignment may become stale as resource availability changes. 
\sys therefore defers binding until all predecessor tokens are satisfied. The operation then enters the ready queue, and the policy selects a \texttt{pCtx} using the current resource state. 
This late-binding rule keeps physical assignments aligned with the resources available at dispatch time.

\heading{Binding and Submission.}
For each selected ready operation $o$, the policy proposes a \texttt{pCtx}. 
Because policies are pluggable and resource state may change between selection and submission, the dispatcher treats the proposal as tentative and validates whether the \texttt{pCtx} remains eligible.
If validation succeeds, it activates or reuses the context's lease, selects a backend stream, submits the immutable operation descriptor, and records the operation as outstanding work.
The binding is committed only after successful submission.
If submission fails, the operation remains ready for rescheduling.

\heading{Completion Propagation.}
Each submitted operation is associated with a backend completion record. 
When completion is observed, the dispatcher satisfies the logical token in the \texttt{vCtx}, potentially making dependent operations ready for dispatch. 
It also decrements the outstanding-work count of the associated \texttt{pCtx}. 
When the count of a draining \texttt{pCtx} reaches zero, the dispatcher releases its lease and transitions it to either \emph{idle} or \emph{reserved}, depending on whether the allocation has been reserved for anticipated work.

\subsection{Policy Interface}
\label{sec:policy-interface}

\sys defines three callbacks implemented by each scheduling policy. 
Let $S$ summarize the current \texttt{pCtx} states, leases, reservations, and dispatch gates.

\begin{myitemize}
\item $\textsc{Pick}(\mathcal{Q},S)\rightarrow o\ \text{or}\ \bot$ selects the next operation from the ready queue $\mathcal{Q}$, or defers dispatch.

\item $\textsc{Select}(o,\mathcal{E},S)\rightarrow p$ selects a candidate \texttt{pCtx} $p\in\mathcal{E}$, where $\mathcal{E}$ is the set of non-conflicting \texttt{pCtxs} that can execute $o$ unchanged.

\item
$\textsc{Prepare}(h,S)\rightarrow R$ produces a resource-preparation plan for a lookahead hint $h$. 
The plan may identify a target allocation, reserve an available \texttt{pCtx}, or close the dispatch gates of conflicting \texttt{pCtxs}.
\end{myitemize}

The policy determines which ready operation to dispatch, where to execute it, and which resources to prepare.
The runtime validates each decision against descriptor compatibility, the logical dependency state, and current leases before changing physical state.

\section{Policy Instantiations}
\label{sec:policies}

We instantiate two simple policies to illustrate how the mechanism supports different scheduling objectives.
Both use the same \textsc{Pick}, \textsc{Select}, and \textsc{Prepare} callbacks and serve only as examples of the interface.

\heading{Performance Profiles.}
\sys profiles each recurring launch configuration at every allocation size exposed by the \texttt{pCtx} pool.
The resulting scaling profile captures the marginal benefit of additional SMs and guides physical allocation without requiring an explicit compute- or memory-bound classification. 
Offline profiles and performance models are widely used in GPU scheduling and resource control~\cite{fastermoe,clock,choi2022serving,sgdrc}.

\heading{Throughput-Oriented Policy.}
The default policy first maximizes concurrency, then greedily assigns remaining SMs to maximize aggregate normalized progress.
Let $\widehat{T}_o(m)$ be the profiled execution time of operation $o$ on $m$ SMs, and define
\[
\begin{aligned}
P_o(m)&=\frac{\widehat{T}_o(m_o^{\min})}{\widehat{T}_o(m)}, \ \ 
G_o(p,p')=\frac{P_o(m(p'))-P_o(m(p))}
{m(p')-m(p)},
\end{aligned}
\]
where $m_o^{\min}$ is the minimum supported allocation, $m(p)$ is the size of \texttt{pCtx} $p$, and $p'$ is a wider alternative.
Among feasible plans assigning the maximum number of ready operations, the policy maximizes
\[
\sum_{o\in\operatorname{dom}(\mathcal{A})}
P_o\bigl(m(\mathcal{A}[o])\bigr).
\]

Algorithm~\ref{alg:tput-policy} greedily approximates this objective.
It first assigns minimum-size, non-overlapping \texttt{pCtxs} to the maximum number of ready operations using age-aware round-robin across \texttt{vCtxs}, then repeatedly applies the feasible positive-gain expansion with the largest $G_o$.
Feasibility requires preserving non-overlap with all other assignments.
Kernels that gain little from additional SMs therefore remain on narrow \texttt{pCtxs}, leaving wider allocations to kernels that continue to scale.
\textsc{Pick} returns the next planned operation, \textsc{Select} returns $\mathcal{A}[o]$, and \textsc{Prepare} is unused.

\begin{algorithm}[t]
\caption{\textsc{Throughput-Oriented} resource assignment.}
\label{alg:tput-policy}
\small
\begin{algorithmic}[1]
\Function{TputPlan}{$\mathcal{Q},S$}
    \State $\mathcal{A} \gets \Call{MinBind}{\mathcal{Q},S}$
    \While{a feasible positive-gain expansion exists}
        \State $(o,p') \gets
        \displaystyle\arg\max_{\text{feasible }(o,p')}
        G_o(\mathcal{A}[o],p')$
        \State $\mathcal{A}[o] \gets p'$
    \EndWhile
    \State \Return $\mathcal{A}$
\EndFunction
\end{algorithmic}
\end{algorithm}

\heading{\textsc{TPOT-First} Policy.}
\textsc{TPOT-First} extends the \textsc{Throughput-Oriented} policy to prioritize decode TPOT.
It maintains an exclusive standing $M_{\mathrm{TPOT}}$-SM \texttt{pCtx}, sized from offline profiles to meet the TPOT target while leaving remaining resources to best-effort work.
For each decode $o$, the policy derives a next-token deadline $d_o$ from the previous token's completion and predicts completion on the reservation, including queued decode work.
If a decode risks missing $d_o$, \textsc{Prepare} drains overlapping best-effort \texttt{pCtxs} before $o$ becomes ready and reserves the smallest wider pre-created \texttt{pCtx} predicted to meet the deadline.
Once ready, \textsc{Pick} prioritizes the earliest-deadline at-risk decode, and \textsc{Select} binds it to the prepared \texttt{pCtx}.
If preparation is incomplete, the policy uses an available allocation or defers dispatch.
All other decisions follow the \textsc{Throughput-Oriented} policy. Under bursts, this TPOT-First choice may delay prefills.

Both policies submit identical immutable descriptors and preserve the same happens-before order $\prec_H$.
They differ only in ordering incomparable ready operations and their physical assignments, so each decision remains within the invariant.
\section{Implementation}
\label{sec:impl}

The \sys prototype comprises approximately 6K lines of C++ and 1K lines of CUDA.
An \texttt{LD\_PRELOAD} library intercepts CUDA launches, stream and event operations, memory operations, and synchronization calls; constructs immutable descriptors and dependencies; and retains CUDA handles and argument values in client processes.
A coordination daemon invokes pluggable policies through the \textsc{Pick}, \textsc{Select}, and \textsc{Prepare} callbacks and returns scheduling decisions without rewriting operation descriptors.
Clients and the daemon exchange compact records through preallocated, lock-free shared-memory rings, using batched processing and idle-only wake-ups to reduce communication overhead.

At initialization, \sys establishes a bounded, globally coordinated \texttt{pCtx} pool.
The daemon maintains its physical-allocation and lease state, while each client realizes the corresponding entries locally using NVIDIA Green Contexts and nonblocking streams.
The daemon arbitrates leases across clients.
A client-side submission thread validates the selected lease before submitting the original descriptor through the corresponding local \texttt{pCtx}.
CUDA events are polled in batches to satisfy logical completion tokens and update \texttt{pCtx} availability.
Pre-creation removes context setup from the dispatch path, while local submission keeps CUDA handles, kernel arguments, and device pointers outside the policy process.
\section{Evaluation}

\subsection{Experimental Setup}
\label{sec:eval-setup}

\heading{Platforms.}
% We evaluate \sys on NVIDIA H20 and H200 GPUs.
End-to-end colocation experiments use H20 GPUs.
Dense training, parameter-efficient fine-tuning, and mixed training--inference use one H20, while distributed training uses eight H20 GPUs.
Determinism validation, mechanism and policy evaluation, and overhead measurements use a single H200.
The decoding and trace-driven serving experiments use SGLang~\cite{sglang2024} with Llama-3.1~\cite{llama3} on an H200, with CUDA Graph replay disabled in all compared configurations to enable per-launch binding.
Within each experiment, all systems use identical hardware, software, inputs, and workloads.

\heading{Determinism Protocol.}
We test the resource-control contract at operator and LLM decoding levels.
All runs use the same $X$, $\{B_k\}$, $\{C_k\}$, $H$, and environment defined in~\autoref{sec:motivation}.
The reshaping treatment changes the structures $\{\widetilde{C}_k\}$ through work repartitioning to emulate resource-dependent structural adaptation.
The \sys treatment holds the descriptors and dependencies fixed while changing only \(\{P_k\}\) under matched interference.
For decoding, we replay an autoregressive state 100 times with temperature-zero greedy decoding and record the LM-head logits, softmax probabilities, and selected token.
SGLang is configured for deterministic execution~\cite{sglang_deterministic}, and exclusive fixed-structure runs are bitwise identical, establishing the prerequisite for the contract.

\begin{figure*}[!t]
    \centering
    \includegraphics[width=\textwidth]{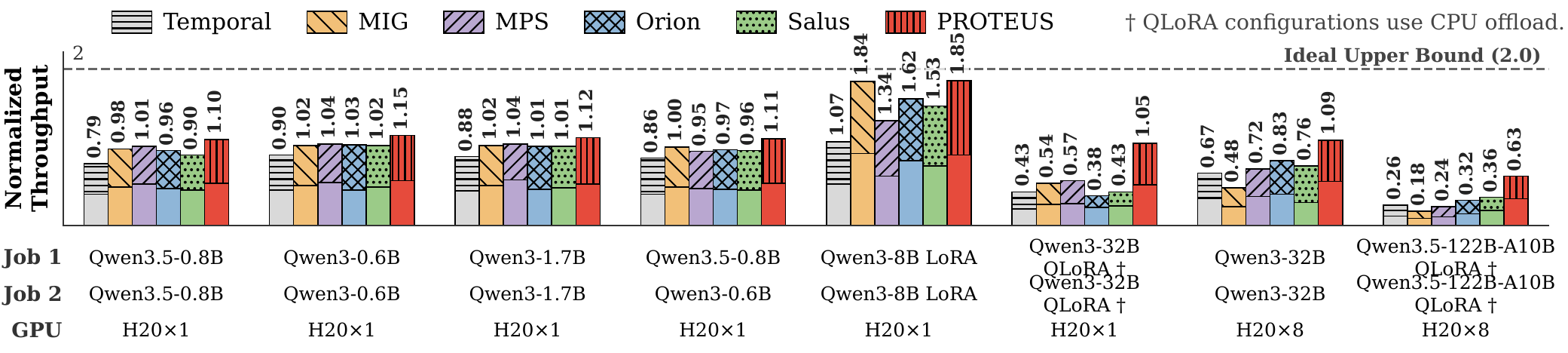}
    \vspace{-2em}
    \caption{Aggregate normalized throughput of colocated training workloads (higher is better).}
    \label{fig:colocated-training}
    \vspace{-0.5em}
\end{figure*}

\begin{figure}[t]
    \centering
    \includegraphics[width=\columnwidth]{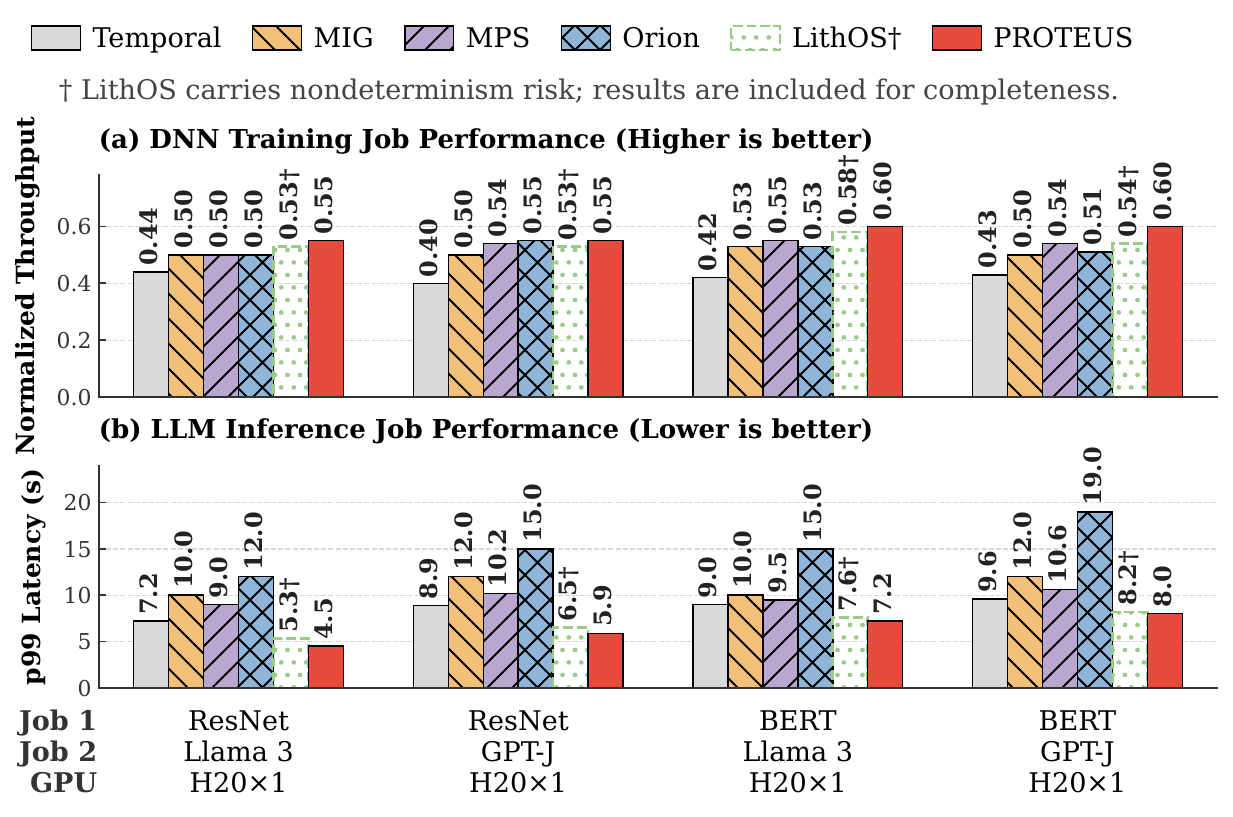}
    \vspace{-1em}
    \caption{Background-training normalized throughput and LLM inference p99 latency under mixed workloads. Each column identifies the training and inference jobs.}
    \label{fig:mixed-workloads}
\end{figure}

\heading{Colocation Workloads and Baselines.}
The colocated-training suite covers dense, LoRA~\cite{hu2022lora}, QLoRA with CPU offloading~\cite{qlora}, and distributed training using Qwen3.5 and Qwen3 models~\cite{qwen35blog,yang2025qwen3technicalreport}.
The mixed suite pairs ResNet and BERT training~\cite{he2016deep,devlin2019bert} with Llama 3 and GPT-J inference~\cite{llama3,wang2021gptj}.
Workload pairs and GPU counts are labeled in \autoref{fig:colocated-training} and \autoref{fig:mixed-workloads}.
For colocated training, we compare \sys with temporal sharing, MPS, MIG, Orion, and Salus~\cite{strati2024orion,yu2020salus}.
Mixed workloads additionally include our reproduction of LithOS~\cite{coppock2025lithos}:
we implement LithOS's atomization logic based on its paper description. While this may not capture all optimizations in the original implementation, our goal is to compare against the \emph{structure-altering} approach.

\heading{Policy Evaluation Setup.}
We evaluate three LLM serving workloads based on Azure LLM Trace 2024~\cite{dynamollm2025}, LongBench~\cite{longbench2024}, and BurstGPT~\cite{burstgpt2025}.
We compare the \textsc{Throughput-Oriented} and \textsc{TPOT-First} policies, which implement the same callbacks with different objectives, while holding the serving stack, model, and workload configuration fixed.
The prefill/decode (P/D)-separated setup identifies the decode instance. The TPOT budget is configured experimentally, and lookahead comes from recorded decodes blocked behind preceding asynchronous memory copies.

% \heading{Metrics.}
% For determinism, we report operator-level maximum absolute differences and bitwise equality, together with decoding-level logit and probability drift and argmax inversions. 
% For colocation, we report normalized throughput and inference p99 latency. 
% For policy evaluation, we report the mean and p50, p90, and p99 TTFT and TPOT, together with SLO violation rates. 
% For overhead, we report p95 end-to-end latency, GPU-memory footprint, and host RSS.

\subsection{Determinism Evaluation}
\label{sec:eval-determinism}

\begin{figure}[t]
    \centering
    \includegraphics[width=\columnwidth]{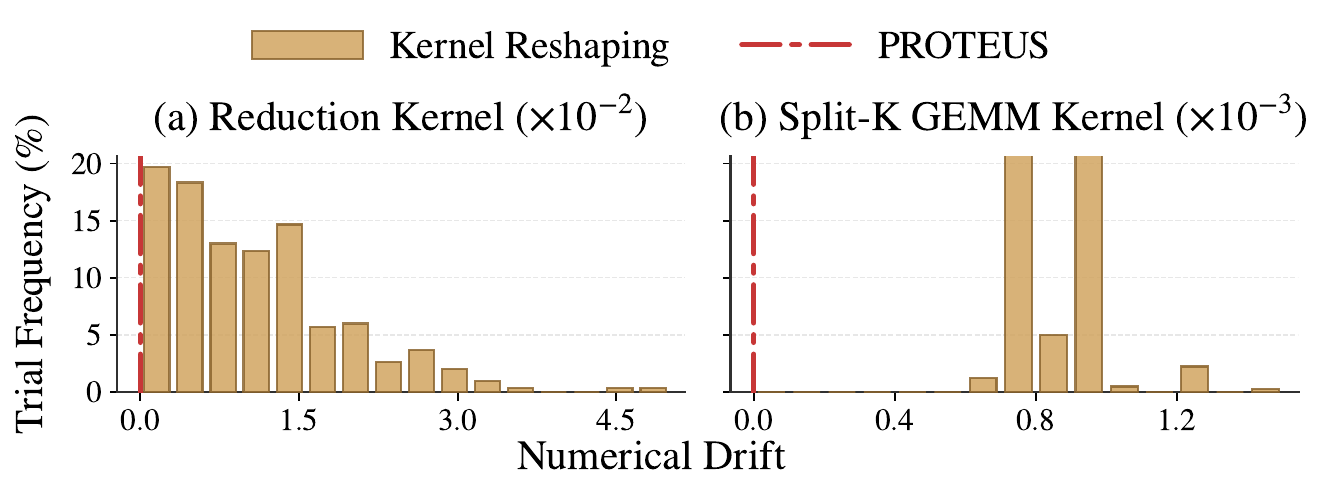}
    \caption{
    Output drift across 1,000 matched-input trials.
    Reshaping changes both operators' outputs, whereas \sys preserves bitwise equality across allocations.
    }
    \label{fig:op-determinism}
\end{figure}

\begin{figure}[!t]
    \centering
    \includegraphics[width=\columnwidth]{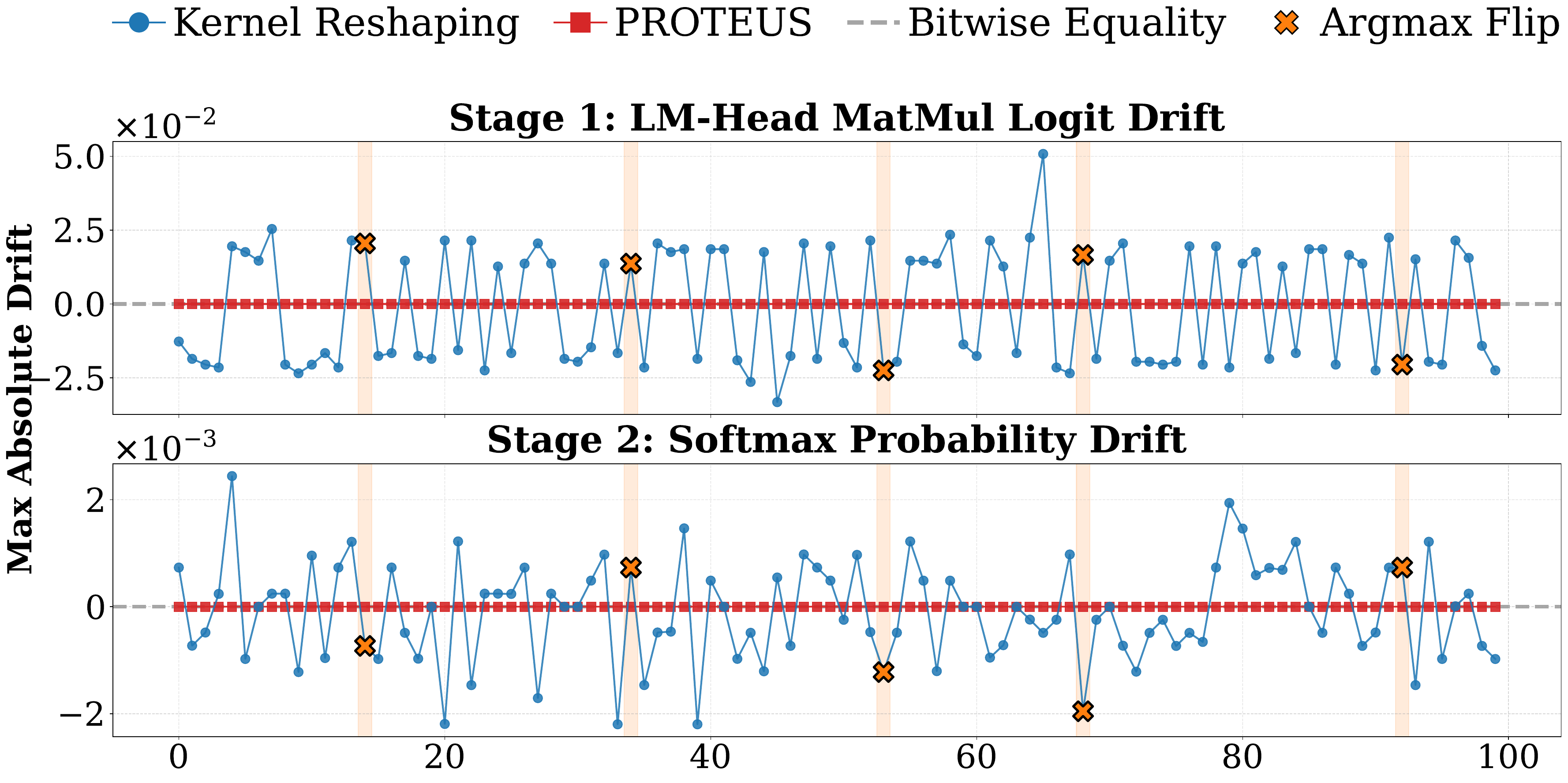}
    \caption{Propagation of reshaping-induced numerical drift through a fixed autoregressive decoding step. Kernel reshaping changes the selected token in five of 100 repetitions.}
    \label{fig:determinism}
\end{figure}

\heading{Operator Drift.}
We evaluate a reduction kernel and a Split-K GEMM over 1,000 matched-input trials.
As shown in \autoref{fig:op-determinism}, resource-driven reshaping alters the executed parallel structure through grid repartitioning or changes in work decomposition, producing nonzero output drift in both kernels.
In contrast, \sys varies only the physical allocation while preserving the submitted structure, yielding bitwise-identical outputs across all trials.
These results isolate resource-dependent changes in floating-point grouping as the cause of the observed variation.

\heading{LLM Decoding-Level Consequence.}
At the decoding level, \autoref{fig:determinism} shows that reshaping-induced LM-head logit drift propagates through softmax and changes the selected token in five of 100 repetitions.
\sys produces bitwise-identical logits and probabilities across all repetitions, with no argmax inversions.
Together with the operator-level results, this experiment shows that resource-dependent changes to parallel structure can become externally visible at decoding.
Parallel-structure-preserving rebinding preserves fixed-structure determinism and satisfies the resource-control contract.

\subsection{End-to-End Performance}
\label{eval:e2e}

We evaluate whether parallel-structure-preserving adaptation also improves end-to-end performance for colocated training and mixed training--inference workloads.
The applicable baselines are described in~\autoref{sec:eval-setup}, and each figure identifies its workload pairs directly.

\subsubsection{Colocated LLM Training}
As shown in \autoref{fig:colocated-training}, \sys achieves the highest aggregate normalized throughput in all eight workloads.
Across these workloads, it achieves 1.38--1.63$\times$ the geometric-mean aggregate normalized throughput of the baselines.
Against the strongest baseline, its gain is 1.08--1.11$\times$ for the dense single-GPU pairs and reaches 1.84$\times$ when offloading or distributed execution creates more variable compute availability.
Across all comparisons, \sys achieves up to 3.50$\times$ the throughput of the lowest-throughput baseline.

\heading{Analysis.}
Dense single-GPU workloads keep the GPU near saturation and leave little transient slack for concurrent progress.
CPU offloading and distributed execution introduce transfer, communication, and synchronization phases that create larger fluctuations in available compute~\cite{dapple, dsp, pipad}.
Per-launch physical rebinding assigns idle SMs to colocated jobs without changing their submitted kernels, producing larger gains in these cases.
The largest distributed workload remains below 1.0 because compute reallocation cannot eliminate its CPU-offload and communication bottlenecks.

\subsubsection{Mixed Training--Inference Colocation}

As shown in \autoref{fig:mixed-workloads}, \sys achieves the lowest inference p99 latency across all four workload pairs while attaining the highest or tied-highest background-training throughput.
Using geometric means across workloads, \sys reduces inference p99 latency by 8.1--58.4\% and achieves 1.05--1.36$\times$ the background-training throughput of the compared systems.
It reduces p99 latency by up to 62.5\% and achieves up to 1.43$\times$ the background-training throughput.
LithOS comes closest to \sys in inference latency, but its kernel atomization changes parallel structure and therefore falls outside the resource-control contract.
\sys achieves lower p99 latency and equal or higher training throughput than LithOS while preserving the parallel-structure invariant.

\heading{Analysis.}
Across both CV--LLM and NLP--LLM colocations, reservation prevents latency-critical inference from waiting behind newly admitted training work. 
When inference requires a wider allocation, lookahead-driven reclamation drains only overlapping \texttt{pCtxs}, allowing training to continue on non-overlapping resources. 
This selective preparation explains why \sys consistently reduces inference tail latency without sacrificing background progress across different training and LLM pairs.

\subsection{Mechanism Effectiveness and Policy Flexibility}
\label{sec:policy-adaptation}

\begin{figure}[!t]
    \centering
    \includegraphics[width=\columnwidth]{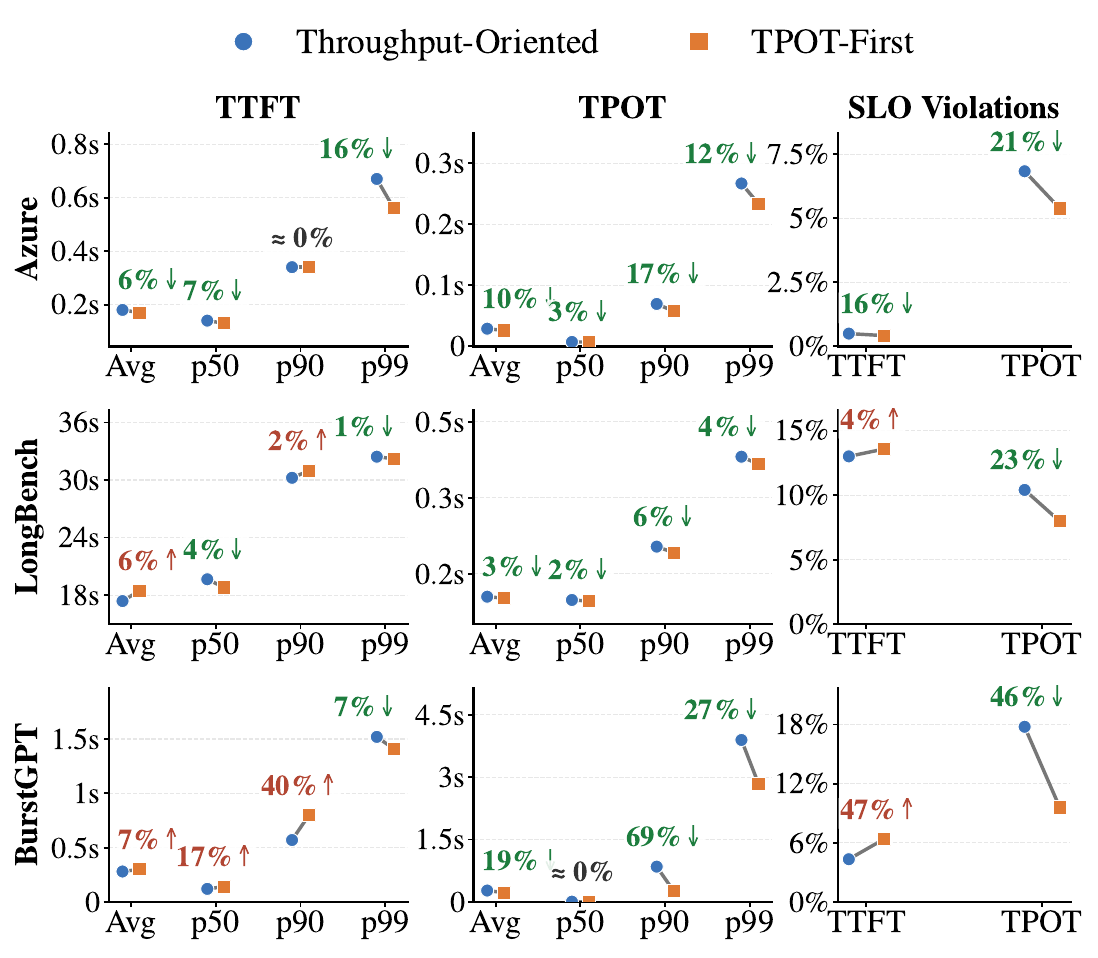}
    \caption{Comparison of the \textsc{Throughput-Oriented} and \textsc{TPOT-First} policies across Azure, LongBench, and BurstGPT.
    Annotations report the relative changes under \textsc{TPOT-First}.}
    \label{fig:policy-comparison}
\end{figure}

As shown in \autoref{fig:policy-comparison}, \textsc{TPOT-First} reduces TPOT SLO violations by 21.2\%, 23.4\%, and 46.1\% on Azure, LongBench, and BurstGPT, respectively. 
The largest improvement occurs on BurstGPT: under its bursty arrivals, p90 TPOT decreases from 847.7\,ms to 262.9\,ms, and p99 TPOT from 3893.0\,ms to 2844.0\,ms. 
The TTFT impact is workload-dependent: the SLO violation rate remains nearly unchanged in absolute terms on Azure, rises slightly on LongBench, and increases from 4.33\% to 6.37\% on BurstGPT.

\heading{Analysis.}
\textsc{TPOT-First} reserves capacity for decode operations and prioritizes deadline-critical decodes. 
When bursts exceed the reservation, lookahead-driven reclamation prepares a wider allocation before the decode becomes ready.
This produces the largest TPOT improvement on BurstGPT and also delays some prefills and increases TTFT violations.
Under Azure and LongBench, the reservation is more often sufficient, yielding TPOT improvements with little or no TTFT penalty.
Together, these results show that policies can pursue distinct scheduling objectives by changing physical assignments and ready-operation order while the parallel-structure invariant remains enforced by the runtime.

\subsection{Overhead Analysis}

\begin{figure}[!t]
    \centering

    \begin{subfigure}[!t]{0.52\columnwidth}
        \centering
        \includegraphics[
            width=\linewidth
        ]{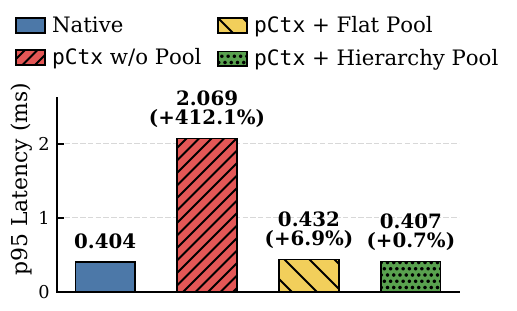}
        \caption{p95 end-to-end latency.}
        \label{fig:overhead:latency}
    \end{subfigure}
    \hfill
    \begin{subfigure}[!t]{0.47\columnwidth}
        \centering
        \includegraphics[
            width=\linewidth
        ]{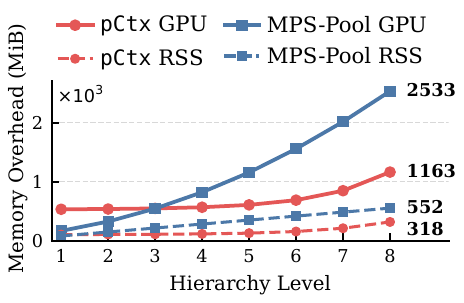}
        \caption{Memory overhead.}
        \label{fig:overhead:memory}
    \end{subfigure}

    \caption{
    Runtime overheads of GreenCtx-based \sys on an H200. (a) p95 latency for a BF16 GEMM with $(M,K,N)=(512,8192,28672)$. (b) GPU memory and host RSS footprints of \textsc{MPS-Pool} and \sys's hierarchical pool.
    }
    \label{fig:overhead}
    \vspace{-0.5em}
\end{figure}

As shown in~\autoref{fig:overhead}, native execution has a p95 end-to-end latency of 404\,\textmu s.
On-demand \texttt{pCtx} creation increases it to 2.069\,ms ($+412.1\%$).
Pre-creation lowers the latency to 432\,\textmu s ($+6.9\%$) with a flat pool and 407\,\textmu s ($+0.7\%$) with the hierarchical pool.
We additionally implement \textsc{MPS-Pool}, a flat baseline that maintains a separately provisioned MPS context for each supported resource share~\cite{nvidia_mps_guide,gslice}.
Despite provisioning a substantially larger context pool, the hierarchical design uses only 1,163\,MiB of GPU memory and 318\,MiB of host RSS at the largest configuration, compared with 2,533\,MiB and 552\,MiB, respectively, for \textsc{MPS-Pool}.
Overall, on an H200, hierarchical pre-creation incurs less than 1\% latency overhead while consuming less than 1\% of GPU memory capacity.

\section{Discussion}

\heading{CUDA Graphs.}
\sys currently disables CUDA Graph replay~\cite{NVIDIA2026CUDAGraphs} and focuses on eager submission, which remains common for dynamic shapes and online serving.
Supporting CUDA Graphs requires context-aware construction or regeneration that preserves dependencies while enabling node-level binding, which we leave to future work.

\heading{Kernel Safety and Fault Isolation.}
The parallel-structure invariant targets deterministic resource adaptation.
Kernel safety and cross-tenant fault containment are orthogonal.
Guardian and KRYPTON provide tenant isolation, while GPUVerify statically checks kernel safety~\cite{Guardian,zhang2025krypton,gpuverify}.
\section{Related Work}

\heading{Determinism.}
Prior work improves reproducibility through numerical algorithms and invariant kernels~\cite{ahrens2020reproducible,zhang2026deterministicinferencetensorparallel}, deterministic programming models~\cite{praun2007implicit,newton2016parallel}, verified compilation and libraries~\cite{boldo2015verified,RepDL}, and runtime enforcement~\cite{qithread,doradd,gond2026llm42enablingdeterminismllm,wild_sdc, opguard}.
Other studies characterize numerical variation in GPU and LLM execution~\cite{test_gpu,xie2025mmasimbitaccuratereferencemodel,yuan2025understanding}.
\sys complements these techniques by preserving parallel structure and application-required ordering under spatial sharing, preventing resource adaptation from introducing structural nondeterminism.

\heading{Spatial Sharing.}
Early work proposed spatial multitasking by partitioning SMs among concurrent applications~\cite{old_gpu_spatial_sharing}.
MPS, Green Contexts, and CU masks expose compute-allocation primitives, while MIG also partitions memory-system resources~\cite{nvidia_mps_guide,nvidia_mig_guide,NVIDIA2025Green,amd_rocm_cu_mask}.
Building on these primitives, MISO, ParvaGPU, and ECLIP enable workload-aware spatial partitioning for GPU colocation~\cite{miso,parvaGPU,eclip}.
Pagoda, GSlice, GPUlets, COLTI, Orion, Laser, and Bless provide finer-grained spatio-temporal scheduling~\cite{pagoda,gslice,choi2022serving,mobin2023colti,strati2024orion,laser,bless}.
Fractional GPUs and SGDRC jointly manage compute and memory-system resources through partitioning and coloring, complementing \sys's compute allocation~\cite{fractional,sgdrc}.
Reshaping-based systems obtain finer control by changing launch geometry or decomposing kernel work~\cite{kernelet,coppock2025lithos,Huang2026ShareNK,hu2026hummingbirdsloorientedgpupreemption}.
In contrast, \sys late-binds immutable launches to physical compute allocations while enforcing the parallel-structure invariant.
\section{Conclusion}
\sys builds GPU resource control around the parallel-structure invariant. It combines virtual contexts, replaceable physical allocations, and validated late binding to adapt each launch's allocation while preserving submitted work, required dependencies, and fixed-structure determinism.

\bibliographystyle{ACM-Reference-Format}
\bibliography{paper}

\end{document}